# First Earth-based Detection of a Superbolide on Jupiter

**Short title: A BOLIDE IN JUPITER ATMOSPHERE**

R. Hueso[1*], A. Wesley[2], C. Go[3], S. Pérez-Hoyos[1], M. H. Wong[4], L. N. Fletcher[5], A. Sánchez-Lavega[1], M. B. E. Boslough[6], I. de Pater[4], G. S. Orton[7], A. A. Simon-Miller[8], S. G. Djorgovski[9], M. L. Edwards[10], H. B. Hammel[11], J. T. Clarke[12], K. S. Noll[13], P. A. Yanamandra-Fisher[7]

[1] Universidad del País Vasco, 48013 Bilbao, Spain

[2] Acquerra Pty. Ltd., 82 Merryville Drive, Murrumbateman, NSW 2582, Australia

[3] Physics Department-University of San Carlos, Cebu City, Philippines

[4] Astronomy Department, 601 Campbell Hall, University of California, Berkeley, CA 94720, USA

[5] Atmospheric, Oceanic and Planetary Physics, Clarendon Laboratory, University of Oxford, Parks Road, Oxford OX1 3PU, UK.

[6] Sandia National Laboratories, Albuquerque, NM 87185, USA

[7] Jet Propulsion Laboratory, California Institute of Technology, Pasadena, CA 91109, USA

[8] Goddard Space Flight Center, Greenbelt, MD 20771, USA

[9] Division of Physics, Mathematics, and Astronomy, California Institute of Technology, Pasadena, CA 91125, USA

[10] Gemini Observatory, c/o AURA Casilla 603, La Serena, Chile

[11]Space Science Institute, 4750 Walnut Avenue, Suite 205, Boulder, CO 80301, USA.

[12] Boston University, Boston, MA 02215, USA.

[13] Space Telescope Science Institute, 3700 San Martin Drive, Baltimore, MD 21218, USA

* To whom correspondence should be addressed: E-mail: ricardo.hueso@ehu.es

**Abstract:** Cosmic collisions on planets cause detectable optical flashes that range from terrestrial shooting stars to bright fireballs. On June 3, 2010 a bolide in Jupiter's atmosphere was simultaneously observed from the Earth by two amateur astronomers observing Jupiter in red and blue wavelengths. The bolide appeared as a flash of 2 s duration in video recording data of the planet. The analysis of the light curve of the observations results in an estimated energy of the impact of $0.9$-$4.0 \times 10^{15}$ J which corresponds to a colliding body of 8-13 m diameter assuming a mean density of 2 g $cm^{-3}$. Images acquired a few days later by the Hubble Space Telescope and other large ground-based facilities did not show any signature of aerosol debris, temperature or chemical composition anomaly, confirming that the body was small and destroyed in Jupiter's upper atmosphere. Several collisions of this size may happen on Jupiter on a yearly basis. A systematic study of the impact rate and size of these bolides can enable an empirical determination of the flux of meteoroids in Jupiter with implications for the populations of small bodies in the outer Solar System and may allow a better quantification of the threat of impacting bodies to Earth. The serendipitous recording of this optical flash opens a new window in the observation of Jupiter with small telescopes.

## 1. INTRODUCTION

The direct observation of collisions on Solar System objects outside the Earth constitutes a rare event that has been directly observed only a few times outside the Earth. Small meteoritic impacts on the Moon have been observed from ground-based observations (Ortiz et al. 2000); a meteoroid entering Mars atmosphere was observed by the Spirit Rover (Bell et al. 2004; Selsis et al. 2004); the Voyager 1 observed a small light flash on Jupiter associated with the entry of a 11 kg meteoroid (Cook et al. 1981) and the Galileo spacecraft observed the bright fireballs produced by the entry of some of the Shoemaker-Levy 9 (SL9) series of impacts in 1994 (Chapman et al. 1996; Martin et al. 1997). Recently, the debris of a large impact on Jupiter comparable to some of the middle range SL9 impacts was observed (Sánchez-Lavega et al. 2010; Hammel et al. 2010) but the direct impact of this object could not be detected since it happened at the night-side of the planet. In this paper we report the observations of a bolide in the upper atmosphere of Jupiter produced by the impact of an unknown object on June 3, 2010. We also present observations targeted to look for the aerosol debris and temperature and chemical anomalies found in previous large-scale jovian impacts; they revealed no detectable signatures in the atmosphere following the impact. The mass and size of the impact body is retrieved from analysis of the light curve of the bolide flash. The results we report support the use of small telescopes for surveying and improving the statistics of unpredicted impacts in the giant planets. Since Jupiter is the most massive planet in the Solar System it constitutes a natural framework to study the atmospheric response to impacts and the nature of the population of small objects able to collide with the planet.

## 2. OBSERVATIONS AND ANALYSIS

The bright flash was simultaneously and independently detected by A. Wesley (AW, Murrumbateman, Australia) and C. Go (CG, Cebu, Phillipines) at 20:31:20 UT on June 3, 2010 using telescopes of 37 cm (AW) and 28 cm (CG) in diameter and fast astronomical cameras at two different wavelengths (red and blue wide passbands with effective wavelengths of 650 and 435 nm) observing at a rate of 60 (AW) and 55 (CG) frames per second (fps) (Figure 1). Both observers used a monochrome Flea3 camera equipped with a ICX618ALA chip. AW was using a red filter from Astrodon, CG was using a blue filter from Edmund Scientific.

Because the flash was detected simultaneously from two different geographical locations, it unambiguously occurred in Jupiter and not in the Earth's atmosphere. The flash occurred close to Jupiter's limb in the equatorial region, at longitude 159ºW (system III) and planetographic latitude 16.5º S. More observations were acquired in the following minutes by the same observers and other amateur astronomers who are contributors to the International Outer Planets Watch survey of Jupiter (Hueso et al. 2010) but they detected neither any immediate remnant of the optical flash, nor any apparent change in the clouds. Follow-up observations were obtained by Hubble Space Telescope (HST) and large ground-based telescopes as will be discussed later.

Since the event was recorded with high temporal resolution it was possible for the first time to obtain from Earth the light curve of a bolide in Jupiter's atmosphere. The optical flash originated from an area smaller than a single pixel and was spread by telescope diffraction and atmospheric turbulence over several pixels of the CCD. Light curves were extracted from both datasets by measuring the excess luminosity in a box of 7x7 pixels, large enough to encompass the entire flash. The photometric signal of the impact area was determined in the previous and later frames without the flash with an uncertainty in the background reference level of 5 Digital Numbers (DN) (6% in brightness) in the red dataset and of 4 DN (4% in brightness) in the blue.

Figure 2A shows the light curves we derived from the two observations. The flash was observable for a total of 2 s. The short duration is similar to the duration of intense flashes on bolides observed on Earth (Brown et al. 2002; Jenniskens et al. 2009). Since both light curves were very similar, a single fit to both datasets was computed to reduce errors in the global behaviour of the light curve. The three main bumps in the combined light curve are consistent with brightness peaks in bolides entering Earth atmosphere. The light curve is not fully symmetrical around the central time. The flash starts smoothly, produces a bright central flash, and then decays faster than the onset.

The flux increase in both light curves points unambiguously to energies of a small object colliding with Jupiter. To calculate the energy, mass and size of the impact body we calibrated the response of the camera and filters used in the observations (see Figure 3) and derived the conversion factor from DNs to $Wm^{-2}$ by scaling the observed and known reflectivity of the Jovian disk (Chanover et al. 1996; Karkoschka, 1998; Hammel et al. 2010) convolved with the system responses. The transformation factors from DNs to $Wm^{-2}$ are 0.14 and 0.044 $Wm^{-2}DN^{-1}$ in the red and blue observations, respectively.

The luminous energies appearing in Figure 2B are computed by multiplying these conversion factors by the excess luminosity in DNs and the total area of the spot used for measurements. Fits to the light curves translate into time integrated luminous energies of $(2.9\pm0.7)\times10^{13}$ and $(1.6\pm0.3)\times10^{13}$ J in the red and blue datasets respectively. Assuming the flash is blackbody emission from ablated meteoroid and shock-heated jovian air the difference in energy between both wavelengths yields a blackbody temperature of $T_{BB} = 4400^{+1000}_{-400} K$. This temperature is in the range of temperatures measured for bolides entering Earth atmosphere (3700±100 K for the 2008 TC$_3$ asteroid; Borovicka and Charvat, 2009) and the optical flashes observed in the energetic impacts of the Shoemaker-Levy 9 fragments (7800±600 K; Chapman et al. 1996). For this range of temperatures only 4-10% of the radiant energy is observable at the wavelengths of the observations. Assuming also isotropic emission and a reflection in the jovian clouds of half of the luminous energy emitted downwards to the planet the total optical energy can be constrained to the range 2.0-8.0x10$^{14}$ J. The ratio between optical energy and the total energy of the impacting body depends on the bolide luminous energy and is constrained from observations of Earth colliding objects (Brown et al. 2002). The luminous efficiency for Earth bolides is given by

$$\eta = 0.12 E_0^{0.115}, \tag{1}$$

where $E_0$ is the optical energy measured in kiloton (1 kiloton=4.185x10$^{12}$ J). Assuming this relation holds true for the present case we obtain η=0.2 which leads to a total energy for the 2010 Jupiter bolide of 1.0-4.0x10$^{15}$ J (equivalent to 250-1000 kiloton). This energy falls in the range of intense superbolides (Ceplecha et al. 1999) that enter Earth's atmosphere and is 5-50 times less energetic than recent estimates of the Tunguska event (3-5MTn; Boslough et al. 2008). Assuming a mean impact velocity with Jupiter of 60 km s$^{-1}$ and a bulk density of 2,000 kg m$^{-3}$ we estimate the mass of the impactor body and its size to be of order 500-2000 Tn and 8-13 m (diameter).

Collisions with Jupiter within this mass range have never been detected before, so the effects on Jupiter's atmosphere provide a fascinating comparison to impacts of the SL9 fragments (Chapman, 1996; Harrington et al. 2004) and the July 2009 Jupiter impact (Sanchez-Lavega et al., 2010; Hammel et al., 2010). In the three days following the optical flash a range of ground-based observatories and the Hubble Space Telescope raced to search for signatures of the collision focussing on wavelengths where debris from previous collisions was instantly identifiable. Table 1 summarizes these observations and figure 4 shows examples of selected images. Gemini-N/NIRI and Keck/NIRC2 surveyed reflected sunlight in strong CH4 absorption bands in the 1.5-2.3 um region but observed no evidence for high-altitude debris. VLT/VISIR, Gemini-S/T-ReCS and IRTF/TEXES searched for thermal perturbations and particular chemistry in the impact site (7-25 μm) but saw no signatures of (i) excess thermal energy; (ii) ammonia gas dredged from the troposphere by the rising fireball; or (iii) stratospheric silicate debris. Finally, Hubble Space Telescope observations with the WFC3 instrument confirmed that neither a visibly-dark debris field nor UV-absorbent aerosols were present over the impact longitude. Each of these phenomena were hallmarks of previous impacts in Jupiter (Hammel et al. 2010; Harrington et al. 2004; de Pater et al. 2010; Fletcher et al. 2010; Orton et al. 2010) and their absence from the 2010 collision confirm that this object was considerably smaller. Furthermore, the impactor did not reach the visible cloud decks at 700 mbar, and had no effect on the thermal structure of the lower stratosphere (10-100 mbar).

## 3. DISCUSSION

Compared to meteorites entering Earth's atmosphere the current object has a mass comparable to the 1994 Marshall Islands fireball (Tagliaferri et al. 1995) and is 100 times less massive than Tunguska but is closer to the latter in terms of energy release. Compared to jovian impactors it lies in an unexplored range of masses between the 2009 July impact (Sánchez-Lavega et al. 2010), which had a mass $10^5$ times larger and produced strong atmospheric effects observable for months (Hammel et al. 2010), and the small fireball observed by Voyager 1 in 1981, $10^5$ times less massive (Cook and Duxbury, 1981).

There are no models for the population of objects of this size range in the outer Solar System. Models of the flux of meteorites on the Earth predict objects of this size to collide with our planet every 6–15 years (Brown et al. 2002). An extrapolation of the expected impact rate in the jovian system from the cratering record in Galilean satellites (Schenk et al., 2004; Zahnle et al. 2003) would predict 1 impact of this kind per year on Jupiter. On the other hand, based upon an extrapolation of the dynamical models of comets and asteroids in orbits prone to Jupiter encounters (Levison et al. 2000) one would expect 30-100 such collisions every year.

The fact that the flash of a body of this size only lasts for 1-2 s indicates that these objects are difficult to detect in occasional observations and require a continuous filming of the planet at a high frame rate. From the strong signal of the detection, the same technique and equipment makes it possible to detect objects 5 times less luminous (diameters on the order of 5 m). Depending on the size distribution for this mass range, these smaller bodies collide with Jupiter 2–5 times more frequently than the current meteorite estimations. Alternatively, objects of slightly larger size (d>15 m) could also be detected in Saturn. Telescopes with a minimum size in the range 15 – 20 cm in diameter equipped with webcams and video recorders provide the best means to calculate the impact rate on Jupiter. Automatic detection of the impacts can be performed with image segmentation and subtraction methods adapted from those used in optical GRB detections and gravitational microlensing (Alard and Lupton, 1998). A careful examination of stored observations of Jupiter by different amateurs covering thousands of hours of video could be a first step toward this effort.

**Note added:** A second bolide on Jupiter was detected by amateur astronomers Masayaki Tachikawa and Aoki Kazuo on 2010 August 20 at 18h22m while this paper was being reviewed. Both observers recorded color video of an optical flash of 1.5 s duration similar in intensity and appearance to the one here reported. No signatures of the impact were detected in near-infrared data obtained with NIRC2 on the Keck telescope ~2 Jupiter rotations after the impact or in amateur observations 1 Jupiter rotation after the impact. The frequency of bolides in Jupiter atmosphere might be closer to the higher limits shown above and high enough to provide observational constrains of the population of bodies of the 10 m size range that collide with Jupiter.

**Acknowledgements**

This work was supported by the Spanish MICIIN project AYA2009-10701 with FEDER and Grupos Gobierno Vasco IT-464-07. LNF was supported by a Glasstone Science Fellowship at the University of Oxford. G.S.O. and P.A.Y.-F. acknowledge support from NASA grants to the Jet Propulsion Laboratory, California Institute of Technology. S.G.D. acknowledges a partial support from the NSF grant AST-0909182, and from the Ajax Foundation. We thank J. Harrington for discussions and A. Stephens, C. Trujillo, J. Radomski, T. Greathouse, M. Richter and C. Tsang for obtaining part of the ground-based observations. This work was partially based on observations from the following telescopes: (1) HST (program GO/DD-12119), with support provided by NASA through a grant from the Space Telescope Science Institute, which is operated by the Association of Universities for Research in Astronomy, Inc., under NASA contract NAS 5-26555.(2) TRECS and NIRI at the Gemini Observatory, which is operated by the Association of Universities for Research in Astronomy, Inc., under agreement with the NSF on behalf of the Gemini partnership: the National Science Foundation (United States), the Science and Technology Facilities Council (United Kingdom), the National Research Council (Canada), CONICYT (Chile), the Australian Research Council (Australia), Ministério da Ciência e Tecnologia (Brazil) and Ministerio de Ciencia, Tecnología e Innovación Productiva (Argentina). (3) VLT/VISIR at the European Organisation for Astronomical Research in the Southern Hemisphere, Chile. (4) NIRC2 at the W.M. Keck Observatory, which is operated as a scientific partnership among the California Institute of Technology, the University of California and the National Aeronautics and Space Administration. The Observatory was made possible by the generous financial support of the W.M. Keck Foundation. (5) TEXES at the Infrared Telescope Facility, which is operated by the University of Hawaii under Cooperative Agreement nº NNX-08AE38A with the National Aeronautics and Space Administration, Science Mission Directorate, Planetary Astronomy Program.

**Table 1**

| Telescope & Observing program | Time (UT) | hr after impact | Wavelengths (μm) | Sensitivity |
|---|---|---|---|---|
| Keck/NIRC2 | 2010-06-04 1500UT | +18.5 | 2.12, 2.17, 1.58, 1.29, 2.13, 2.06 | A &C |
| IRTF/TEXES | 2010-06-04 1500UT | +18.5 | Mid IR: 10.34, 10.74 | T & N |
| Gemini N/NIRI GN-2010A-DD-4 | 2010-06-04 1500UT | +18.5 | Near IR: 1.69, 2.11, 2.17 | A&C |
| Gemini S/T-ReCS GS-2010A-DD-6 | 2010-06-05 1030UT | +38.0 | Mid IR: 7.9, 8.8, 10.4, 18.3 | T, N, A |
| ESO-VLT /VISIR 60.A-9800(I) | 2010-06-05 1030UT | +38.0 | Mid IR: 7.9, 8.7, 10.7, 13.04, 17.54,18.65, 19.50 | T, N, A |
| HST – WFC3 GO/DD 12119 | 2010-06-07 1030UT | +77.5 | UV: 0.225, 0.275, 0.343 Visible: 0.395, 0.502, 0.631, 0.275 Near IR: 0.889, 0.953 | A C A & C |

(C) Clouds; (A) Upper atmosphere aerosols; (T) Thermal imaging; (N) $NH_3$ and other chemistry modifications. UT times show the approximate time for the central meridian crossing of the impact longitude.

**Figure Legends**

**Figure 1 | Bolide in Jupiter's atmosphere.**

(a) Color composite of Jupiter observations by A.W. at 20:31. Each color channel is built by stacking all frames in a 60 s interval. The flash was added to the color image from the red frames with the optical flash. (b) Bolide flash evolution in red wavelengths as obtained by A.W. Each image is a stack of 10 frames obtained sequentially with a total exposition time of 0.17 s (c) Bolide flash evolution in blue wavelengths as obtained by C.G. Each image is a stack of 5 frames obtained sequentially with a total exposure time of 0.09 s. All times are referenced to the time of the peak of the maximum in the light curve evolution.

**Figure 2 | Flash light curves.**

(a) Flash light curves in Digital Numbers as measured in each frame in both datasets. Vertical lines are the data with their error bars. Vertical continuous red lines are for the red data, vertical dashed blue lines are for the blue data. The data are not scaled. Differences in cloud albedo, filter response and exposition time result in a common background reference level of 92.2±0.5 DNs shown as an horizontal dashed blue line. A global running-averages fit to both datasets is shown with its global uncertainty in the shadowed area. (b) Calibrated light curves for the amount of energy emitted in each wavelength range. Continuous lines are running-average fits to the data.

**Figure 3 | Image calibration.**

Jupiter geometric albedo (black line) and responses of the blue (left dashed line) and red (right dashed line) filters combined with the camera response. Effective wavelengths for each filter are computed as 435 nm (blue data, CG) and 650 nm (red data, AW) where the Jupiter mean geometric albedo is 0.4 and 0.5 respectively.

**Figure 4 | High-resolution observations of the impact location.**

Series of Jupiter images spanning a wide range of wavelengths obtained in the three days following the observed bolide: (a) HST color composite from observations in the visible range on 7 June 2010; (b) HST ultraviolet image at 225 nm also obtained on 7 June 2010; (c) Gemini N/NIRI observations in a near infrared strong methane absorption band (2.12 µm) on 4 June 2010, (d) VLT/VISIR image at 10.5 µm sensitive to Jupiter's temperature, aerosol and composition (specifically ammonia) acquired on 5 June 2010. Debris from a larger impactor should appear dark and well contrasted on (a) and (b) and bright on (c) and (d). Compared to previous Jupiter impact events, a large impactor would also produce a signature in the mid-infrared (7-25 µm) from (i) thermal energy deposited in the atmosphere; (ii) emission from stratospheric particulates, and (iii) emission from ammonia gas dredged from the troposphere into the higher atmosphere. Boxes show the bolide location. Note the absence of signatures of an impact field at any of these wavelengths.

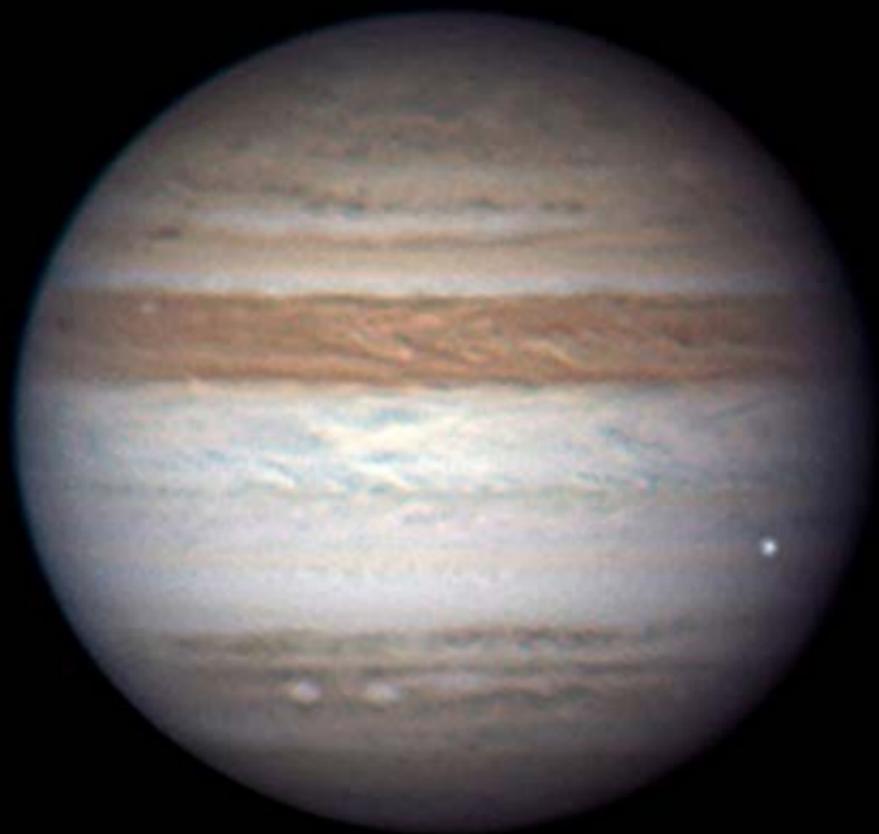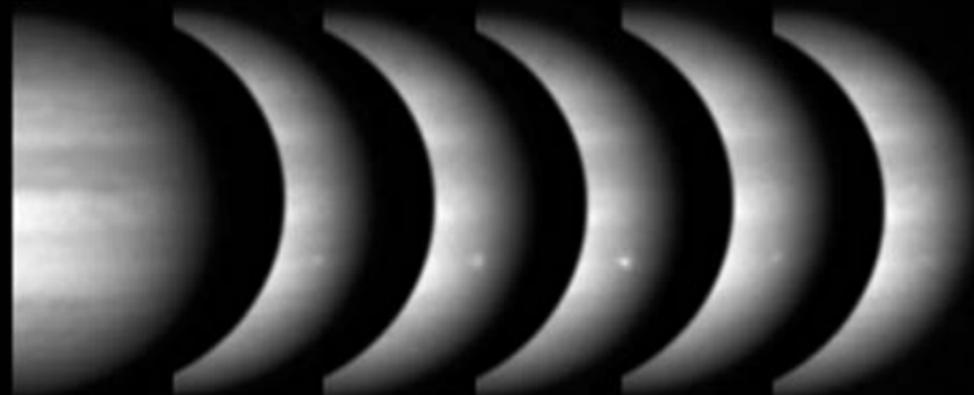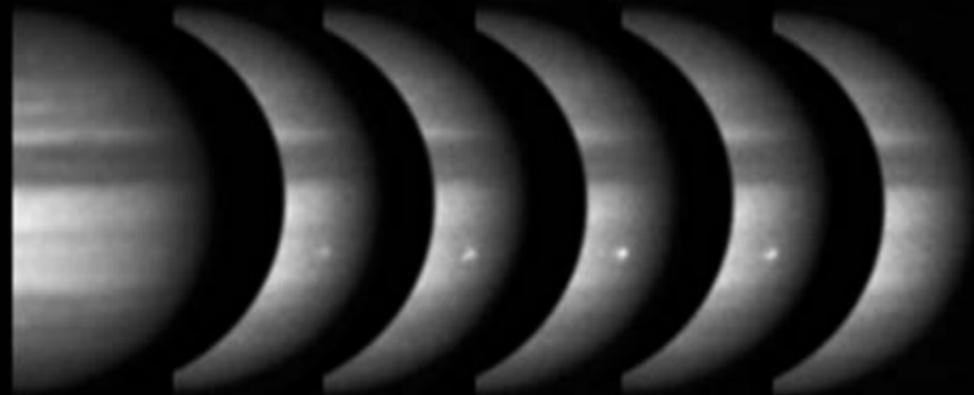

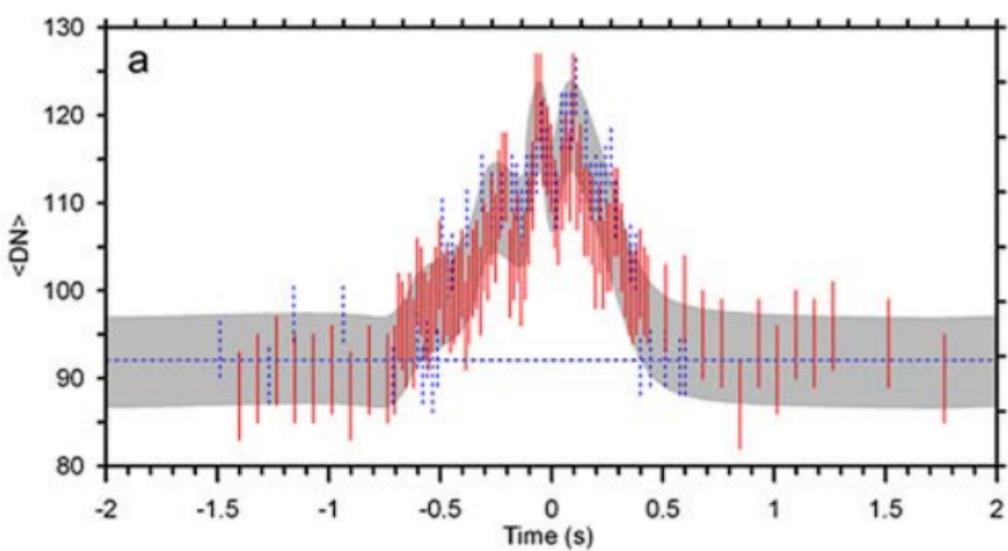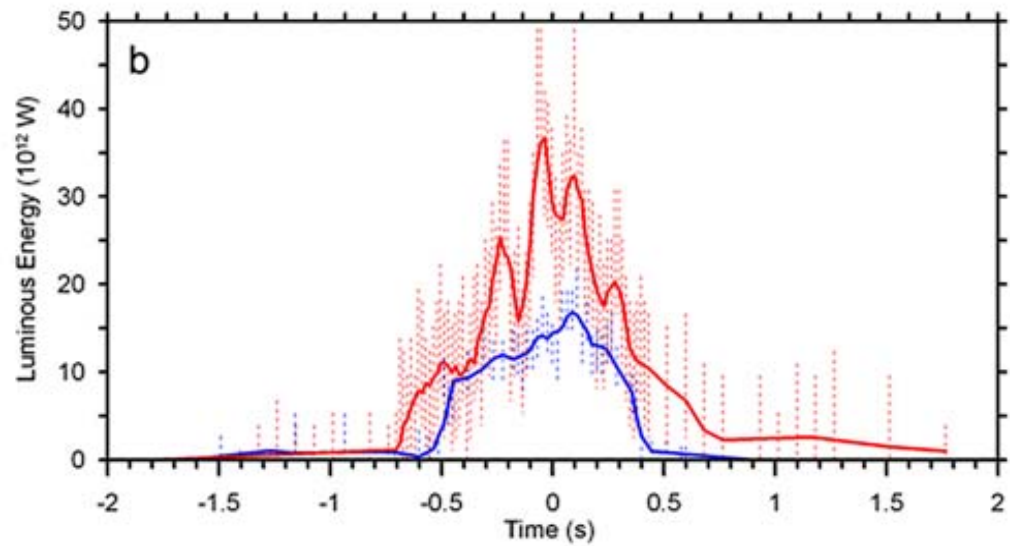

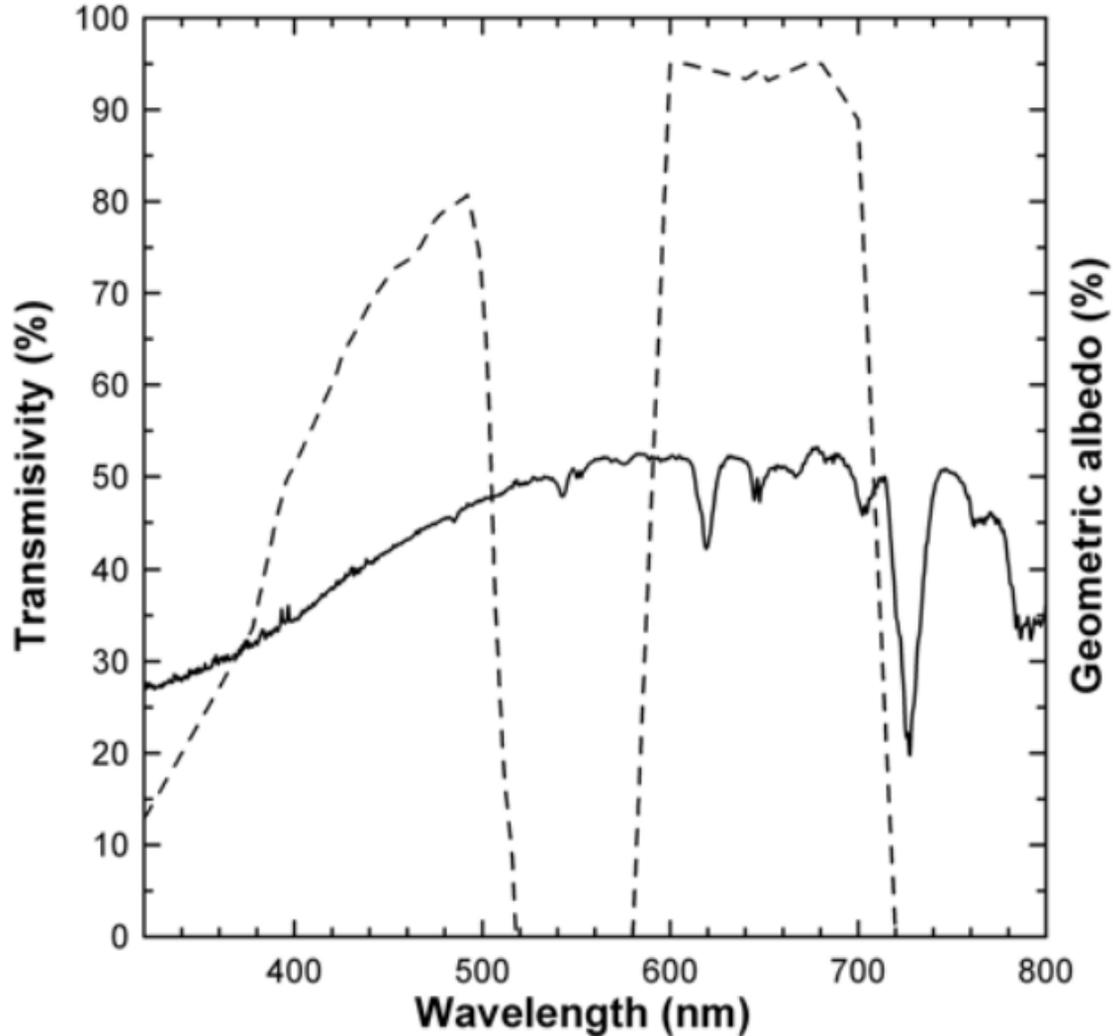

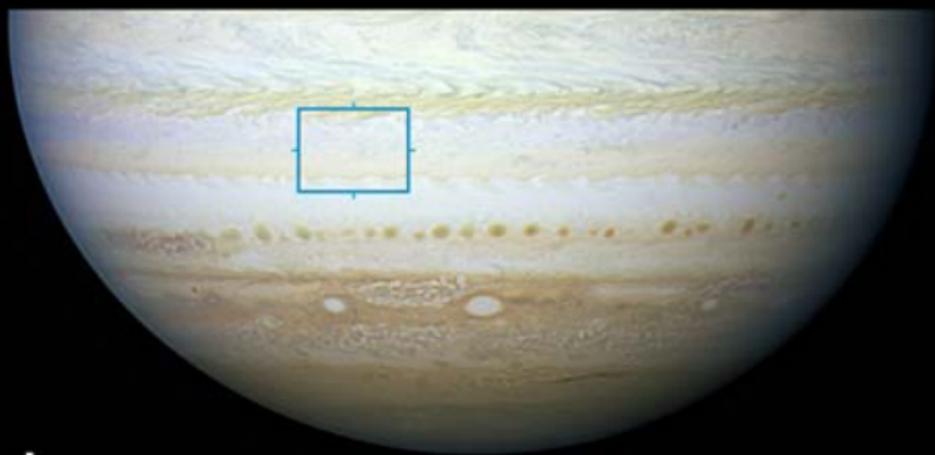
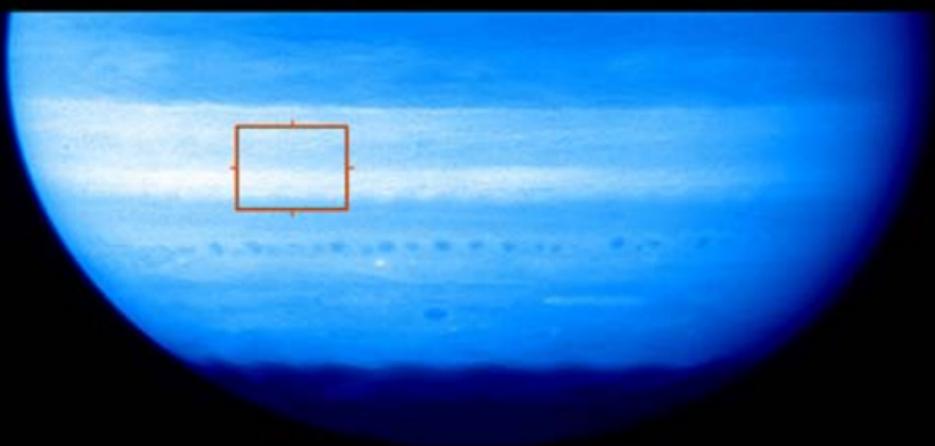
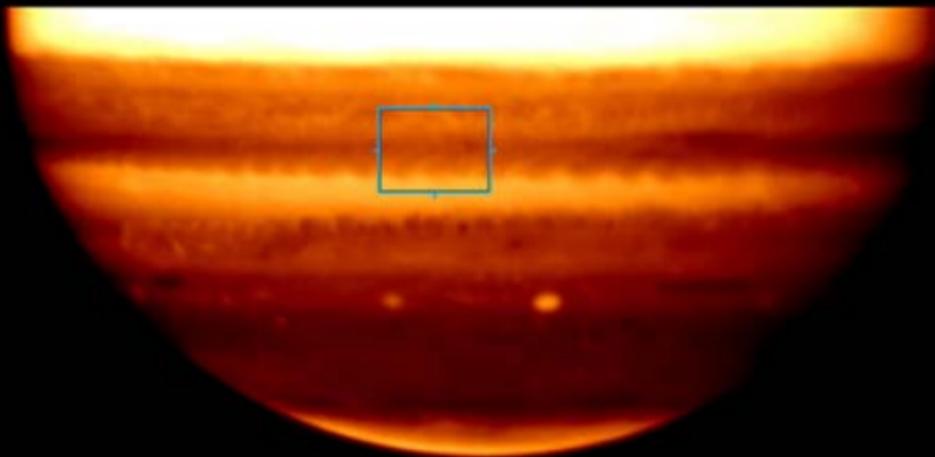
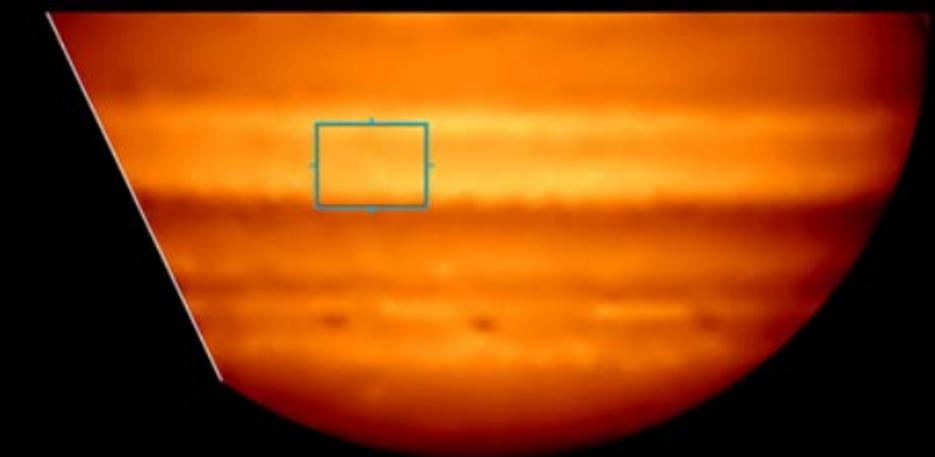